\documentclass[bibnotes,twocolumn,showpacs,amsmath,amssymb,nobalancelastpage,dvips]{revtex4}
\usepackage{bm}
\usepackage{graphics,psfig,epsfig,subfigure,afterpage}
\usepackage{amsfonts}
\usepackage{psfrag}

\newcommand{\be}{\begin{equation}}
\newcommand{\ee}{\end{equation}}
\newcommand{\bea}{\begin{eqnarray}}
\newcommand{\eea}{\end{eqnarray}}
\newcommand{\gdot}{\dot{\gamma}}
\newcommand{\gdotbar}{\bar{\dot{\gamma}}}

\newcommand{\versus}{{\it vs.\/}}

\newcommand{\bw}{\begin{widetext}}
\newcommand{\ew}{\end{widetext}}

\newcommand{\tmax}{t_{\rm max}}
\newcommand{\Tband}{T_{\rm b}}
\newcommand{\Nbase}{N_{\rm base}}

\newcommand{\ommax}{\omega_{\rm max}}

\newcommand{\vecv}[1]{\mathbf{{#1}}}
\newcommand{\tens}[1]{\mathbf{{#1}}}
\newcommand{\nablu}{{\bf \nabla}}

\begin{document}

\title{Linear instability of planar shear banded flow}
\author{S. M. Fielding}
\email{s.m.fielding@leeds.ac.uk}
\affiliation{Polymer IRC and
  School of Physics \& Astronomy, University of Leeds, Leeds LS2
  9JT, United Kingdom} 

\date{\today} 
\begin{abstract}
  We study the linear stability of planar shear banded flow with
  respect to perturbations with wavevector in the plane of the banding
  interface, within the non local Johnson Segalman model.  We find
  that perturbations grow in time, over a range of wavevectors,
  rendering the interface linearly unstable. Results for the unstable
  eigenfunction are used to discuss the nature of the instability. We
  also comment on the stability of phase separated domains to shear
  flow in model H.

\end{abstract}
\maketitle



\begin{figure}[t]
\includegraphics[scale=0.3]{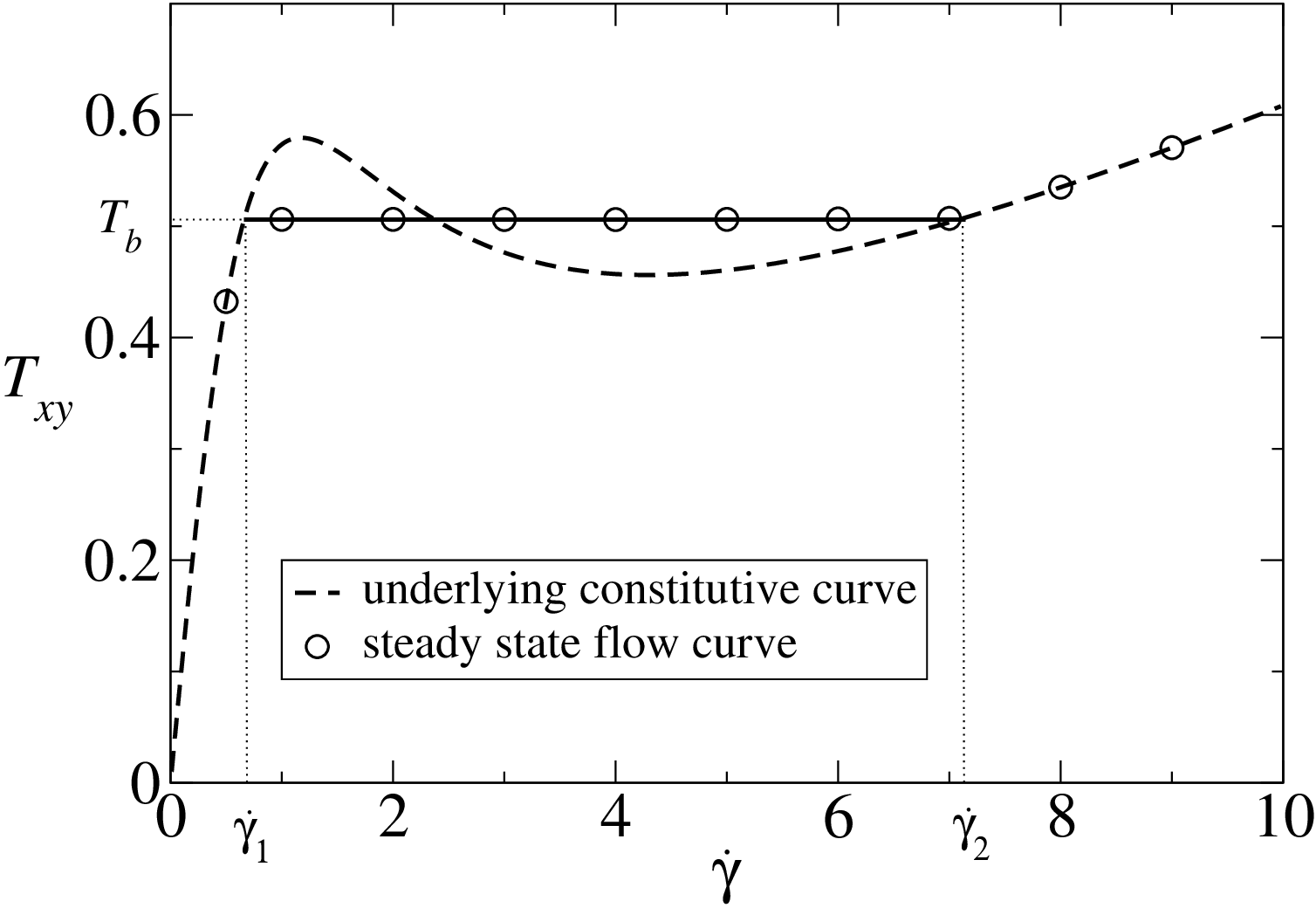}
\caption{Underlying constitutive curve; steady state flow curve. $a=0.3$, $\eta=0.05$. Banding occurs on the plateau. 
\label{fig:flowCurve} } 
\end{figure}

\begin{figure}[t]
  \vspace{-0.5cm} 
\includegraphics[scale=0.3]{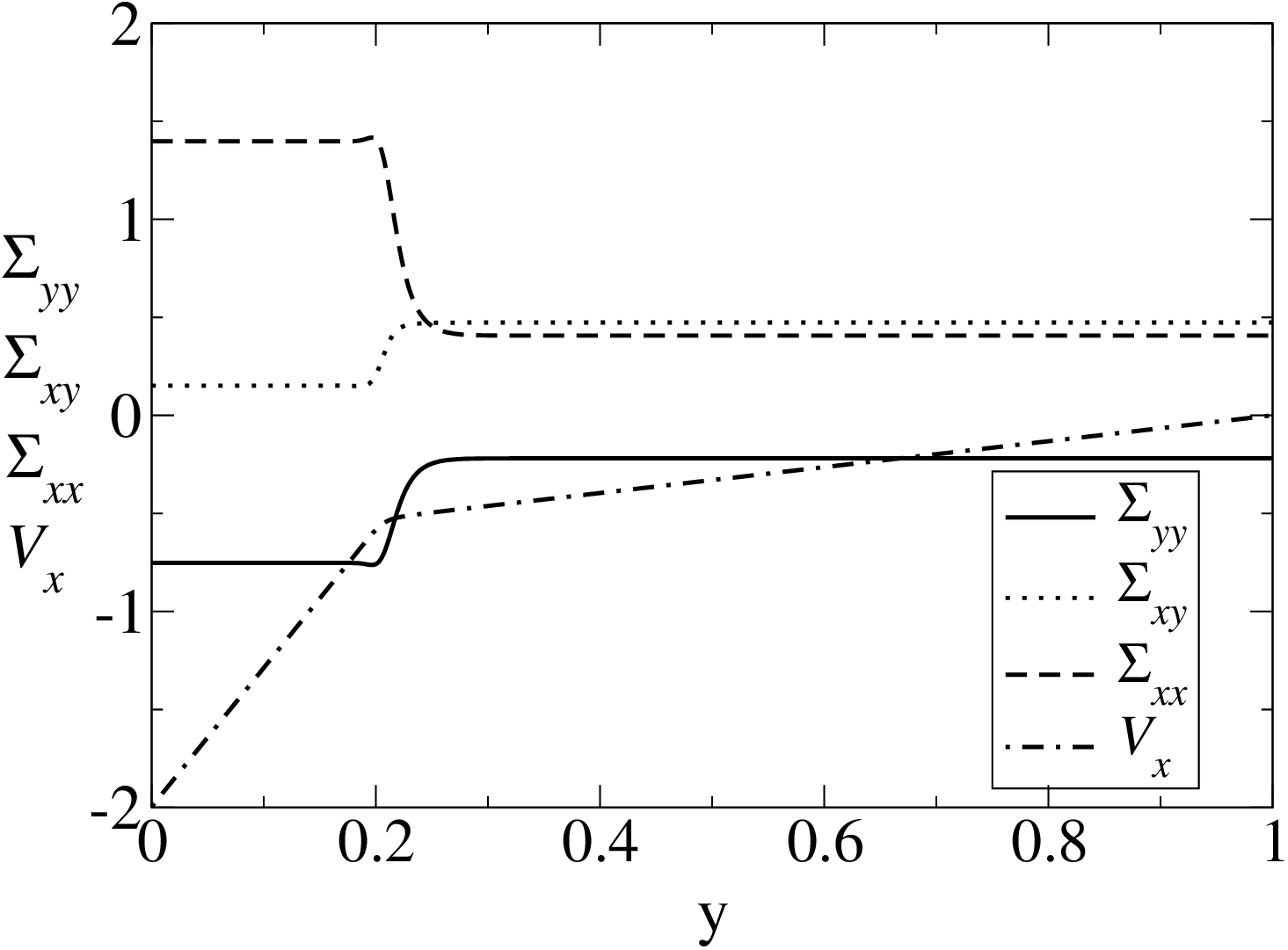}
\caption{1D banded profile, with spatial gradients restricted  to the flow gradient direction, $y$. $\gdotbar=2.0$, towards the left of the plateau in Fig.~\ref{fig:flowCurve}. $l=0.01$, $\Nbase=800$.
\label{fig:profile} } 
\end{figure}

Complex fluids such as wormlike micellar
surfactants~\cite{BritCall97}, lamellar onion phases~\cite{diat93},
polymer solutions~\cite{HilVla02} and soft
glasses~\cite{RayMouBauBerGuiCou02} commonly undergo flow
instabilities and flow-induced transitions that result in spatially
heterogeneous ``shear banded'' states. This effect is captured by
several notable rheological
models~\cite{doiedwards}
in which the underlying constitutive curve of shear stress \versus\ 
shear rate, $T_{xy}(\gdot)$, is non-monotonic
(Fig.~\ref{fig:flowCurve}), allowing the coexistence of bands of
differing shear rate at common shear stress, Fig.~\ref{fig:profile}.
However, most theoretical studies have considered only one spatial
dimension (1D)~\cite{spenley93,lu99},
normal to the interface between the bands (the flow gradient
direction, $y$).  The stability of 1D banded profiles in higher
dimensions has been implicitly assumed, but is in fact an open
question.  In this Letter, therefore, we study numerically the linear
stability of 1D planar shear banded profiles with respect to
perturbations with wavevectors in the  interfacial
plane $(x,z) = ({\rm flow},{\rm vorticity})$.

We work within the Johnson Segalman (JS) model~\cite{johnson77},
modified to include non local diffusive terms~\cite{olmsted99a}. These
account for gradients in the order parameters across the banding
interface, conferring a surface tension. This ``dJS'' model is often
taken as a paradigm of shear banding systems. Our main result will be
that interfacial fluctuations typically grow in time, rendering the 1D
banded profile linearly unstable.  This potentially opens the way to
non trivial interfacial dynamics and could form a starting point for
understanding an emerging body of data revealing erratic fluctuations
of shear banded flows~\cite{WunColLenArnRou01}.
This work is a timely counterpart to new techniques for measuring
interfacial dynamics~\cite{ManSalCol04}.
It is also relevant industrially, to processing instability and oil
extraction.

The model is defined as follows. The generalised Navier Stokes
equation for a viscoelastic material in a Newtonian solvent of
viscosity $\eta$ and density $\rho$ is:
\be
\label{eqn:NS}
\rho(\partial_t + \vecv{V}.\nablu)\vecv{V} = \nablu .(\tens{\Sigma} +
\eta\nablu \vecv{V} -P\tens{I}), \ee
where $\vecv{V}(\vecv{R})$ is the velocity field and
$\vecv{\Sigma}(\vecv{R})$ the viscoelastic part of the stress. For
homogeneous planar shear, $\vecv{V}=y\gdot\vecv{\hat{x}}$, the total
shear stress $T_{xy}=\Sigma_{xy}(\gdot)+\eta\gdot$. The pressure $P$
is determined by incompressibility,
\be
\nablu.\vecv{V}=0.
\label{eqn:incomp}
\ee
The viscoelastic stress evolves with dJS
dynamics~\cite{johnson77,olmsted99a}
\be
\label{eqn:dJS}
 \stackrel{\diamondsuit}{\tens{\Sigma}} = 2 G\tens{D}-\frac{\tens{\Sigma}}{\tau}+ \frac{l^2}{\tau }\nablu^2 \tens{\Sigma},
\ee
with plateau modulus $G$ and relaxation time $\tau$.  The non local
diffusive term accounts for spatial gradients across the interface
between the bands. It arises naturally in models of liquid crystals,
and diffusion of strained polymer molecules~\cite{elkareh89}.
The time derivative
\be 
\label{eqn:tderiv}
\stackrel{\diamondsuit}{\tens{\Sigma}} = (\partial_t +\vecv{V}\cdot\nablu )\,\tens{\Sigma} -
a(\tens{D}\cdot\tens{\Sigma}+\tens{\Sigma}\cdot\tens{D}) - (\tens{\Sigma}\cdot\tens{\Omega} - \tens{\Omega}\cdot\tens{\Sigma}), \nonumber
\ee
in which $\tens{D}$ and $\tens{\Omega}$ are the symmetric and
antisymmetric parts of the velocity gradient tensor, $(\nablu
\vecv{V})_{\alpha\beta}\equiv \partial_\alpha v_\beta$. The ``slip
parameter'' $a$ measures the non-affinity of deformation of the
viscoelastic component~\cite{johnson77}. Slip occurs for $|a|<1$.  The
underlying constitutive curve $T_{xy}(\gdot)$ is then capable of the
non-monotonic behaviour of Fig.~\ref{fig:flowCurve}.

Within this model we consider planar shear between infinite, flat
parallel plates at $y=0,L$. We use units in which $G=1$, $\tau=1$ and
$L=1$; and boundary conditions at $y=0,1$ of
$\partial_y\Sigma_{\alpha\beta}=0\;\forall\;\alpha,\beta$ for the
viscoelastic stress, with no slip and no penetration for the velocity.

For an imposed shear rate $\gdotbar$ in the region of decreasing
stress, $d T_{xy}/d\gdot<0$, homogeneous flow is
unstable~\cite{Yerushalmi70}.  A 1D analysis in the flow gradient
dimension then predicts a separation into two bands of differing shear
rates $\gdot_1, \gdot_2$ at common shear stress, $\Tband$, separated
by an interface of width $O(l)$.  As the applied shear rate $\gdotbar$
is tracked across the banding regime, the relative width-fraction of
the bands adjusts to maintain the constraint $\int dy \gdot(y) =
\gdotbar$, while $\gdot_1, \gdot_2$ and $\Tband$ stay constant,
leading to a plateau in the steady state flow curve
(Fig.~\ref{fig:flowCurve}).


We verified this 1D scenario by numerically evolving
Eqns.~\ref{eqn:NS} to~\ref{eqn:dJS}, allowing spatial variations only
in the flow-gradient direction $y$.  We used a Crank Nicholson
algorithm~\cite{numericalrecipes} within a finite difference scheme on
a uniform mesh of ``full'' points $y_0, y_1 ...y_ {\Nbase}$ for
$\tens{\Sigma}$ and staggered ``half'' points
$y_{\frac{1}{2}},y_{\frac{3}{2}} ...  y_{\Nbase-\frac{1}{2}}$ for
$\vecv{V}$. We evolved with time-step $Dt$ for a time $t_{\rm max}$ to
steady state, checking for convergence to the limit $\Nbase\to\infty,
Dt\to 0, \tmax\to\infty$.

The resulting flow curve is shown in Fig.~\ref{fig:flowCurve}.  A
typical steady state shear banded profile $\vecv{V}(y)$,
$\tens{\Sigma}(y)$ is given in Fig.~\ref{fig:profile}. The velocity
normal to the interface $V_y=0$ in this 1D profile. The smooth
variation of the order parameters across the interface results from
the diffusive term in Eqn.~\ref{eqn:dJS}, which confers an interface
width $O(l)$.  This is in contrast to local models ($l=0$) in which
the interface is a sharp discontinuity. In fact, local models are
pathological in the sense that the banded state is not uniquely
selected, but depends on flow history~\cite{lu99,olmsted99a}.

The linear stability of the sharply banded profiles of local models
was studied by previous authors. Renardy~\cite{renardy} found
instability with respect to interfacial fluctuations of high
wavevector, $q_x\to\infty$, in the local JS model restricted to the
case of a thin high shear band. McLeish~\cite{Mcleish87} studied
capillary flow, for general band thickness.  He demonstrated a long
wavelength ($q_x\to 0$) instability due to the jump in normal stresses
across the interface.  This mechanism was also discussed in
Ref.~\cite{hinch}.

Here we study numerically the {\em non local} case, in which the 1D
banded profile is uniquely selected~\cite{lu99,olmsted99a}. The non
zero interfacial width, $l\ll L$, now confers a surface tension, which
was absent from the local case.  We study general band thicknesses and
the full $({\rm velocity}, {\rm vorticity})$ plane of perturbation
wavevectors $(q_x, q_z)$.

We linearised the model equations~\ref{eqn:NS} to~\ref{eqn:dJS} for
small perturbations (lower case) about the (upper case) base profile,
$\vecv{\tilde{\Phi}}(x, y, z, t)=\vecv{\Phi} (y)+
\vecv{\phi}_{\vecv{q}}(y)\exp(\omega_{\vecv{q}}t+ iq_x x+iq_z z)$.
The vector $\vecv{\Phi}$ comprises all components
$\vecv{\Phi}=(\Sigma_{\alpha\beta}, V_\alpha)$, the pressure being
eliminated by incompressibility. This linearisation results in an
eigenvalue equation with an operator $\mathcal{L}$, which acts linearly
on the perturbation $\vecv{\phi}_{\vecv{q}}(y)$:
\be
\label{eqn:eigen}
\omega_{\vecv{q}}\vecv{\phi}_{\vecv{q}}(y)=\mathcal{L}({\vecv{\Phi}(y), {\vecv{q}}}, \partial_y, \partial_y^2...)\vecv{\phi}_{\vecv{q}}(y).
\ee
For numerical study, we discretized this equation on a staggered mesh.
The 1D base profile $\vecv{\Phi}(y)$ was read in from the calculation
already described.  For narrow interfaces, its uniform mesh had too
many nodes for use in the eigenvalue problem,
so we adapted it to put most attention near the interface. We then
used a NAG routine~\cite{NAG} to find the eigenmodes of this
discretized problem.

The results, discussed below, were checked as follows: (i) for
convergence with respect to mesh structure; (ii) that for a
homogeneous base state on the underlying constitutive curve our
results match those of Ref.~\cite{FieOlm03}; (iii) that for $a=0$,
$l=0$ (the local Oldroyd B model), our method gives results consistent
with Fig. 3 of Ref.~\cite{WilRenRen99}; (iv) that linearisation about
a semi-evolved (non-steady) banded state using the {\em analytically}
derived Eqn.~\ref{eqn:eigen} gives the same results in the limit
$q_x=0, q_z\to 0$ as a particular direct {\em numerical} linearisation
performed about the same profile in the code that evolves the 1D base
state; (v) for robustness with respect to first evolving the base
state on either a uniform or adapted grid, using either a
semi-implicit or explicit algorithm; (vi) that two different methods
of eliminating the pressure (using the Oseen tensor, and the curl
operator) agree.

\begin{figure}[t]
\includegraphics[scale=0.3]{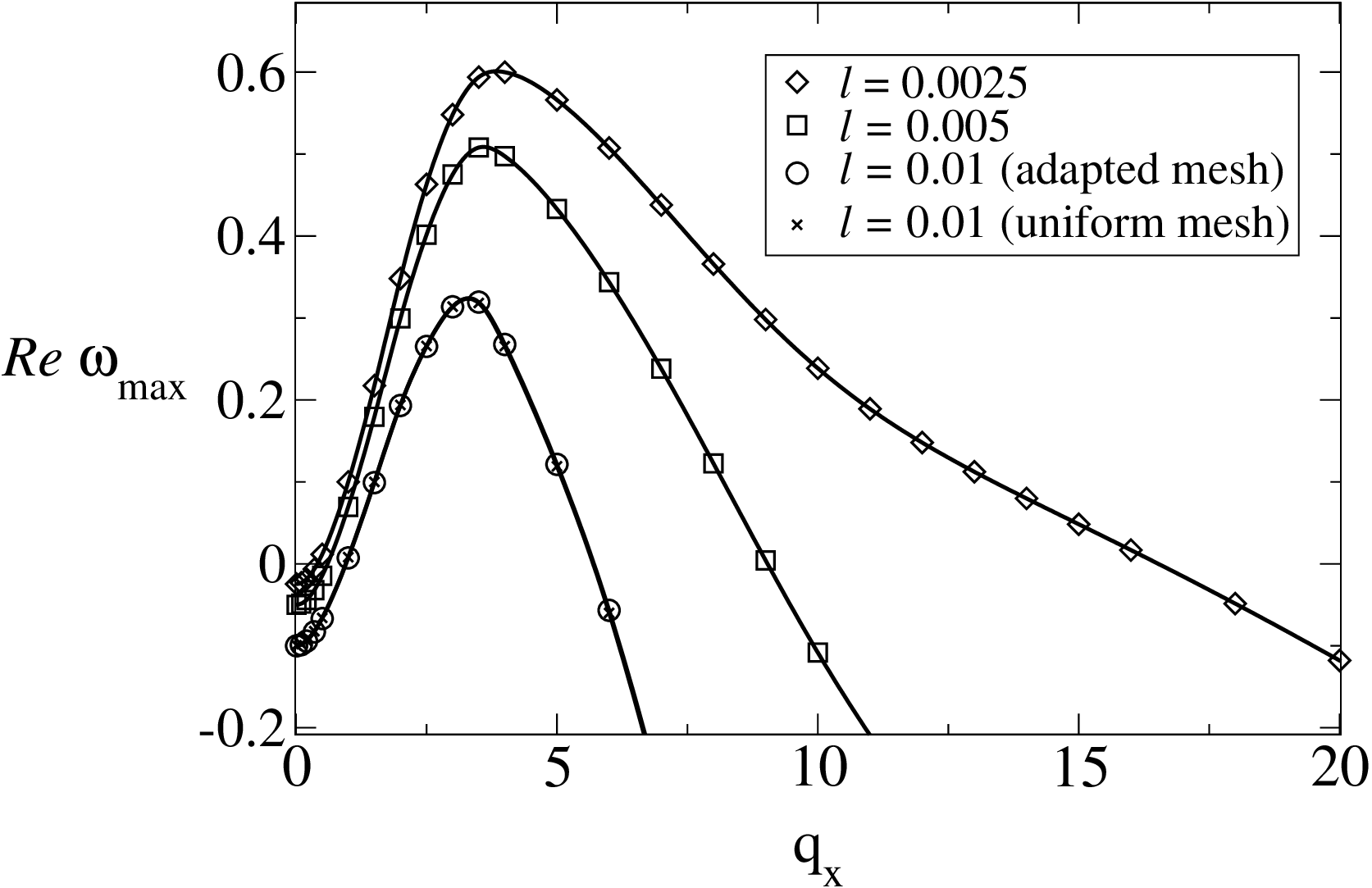}
\caption{Real part of the eigenvalue of the most unstable mode. $a=0.3$,  $\eta=0.05$,  $\gdotbar=2.0$, Reynolds number $\rho/\eta=0$. The data for $l=0.01$ correspond to the base profile in Fig.~\ref{fig:profile}. Symbols: data. Solid lines: cubic splines.
\label{fig:maxEigen} } 
\end{figure}
\begin{figure}[t]
\includegraphics[scale=0.3]{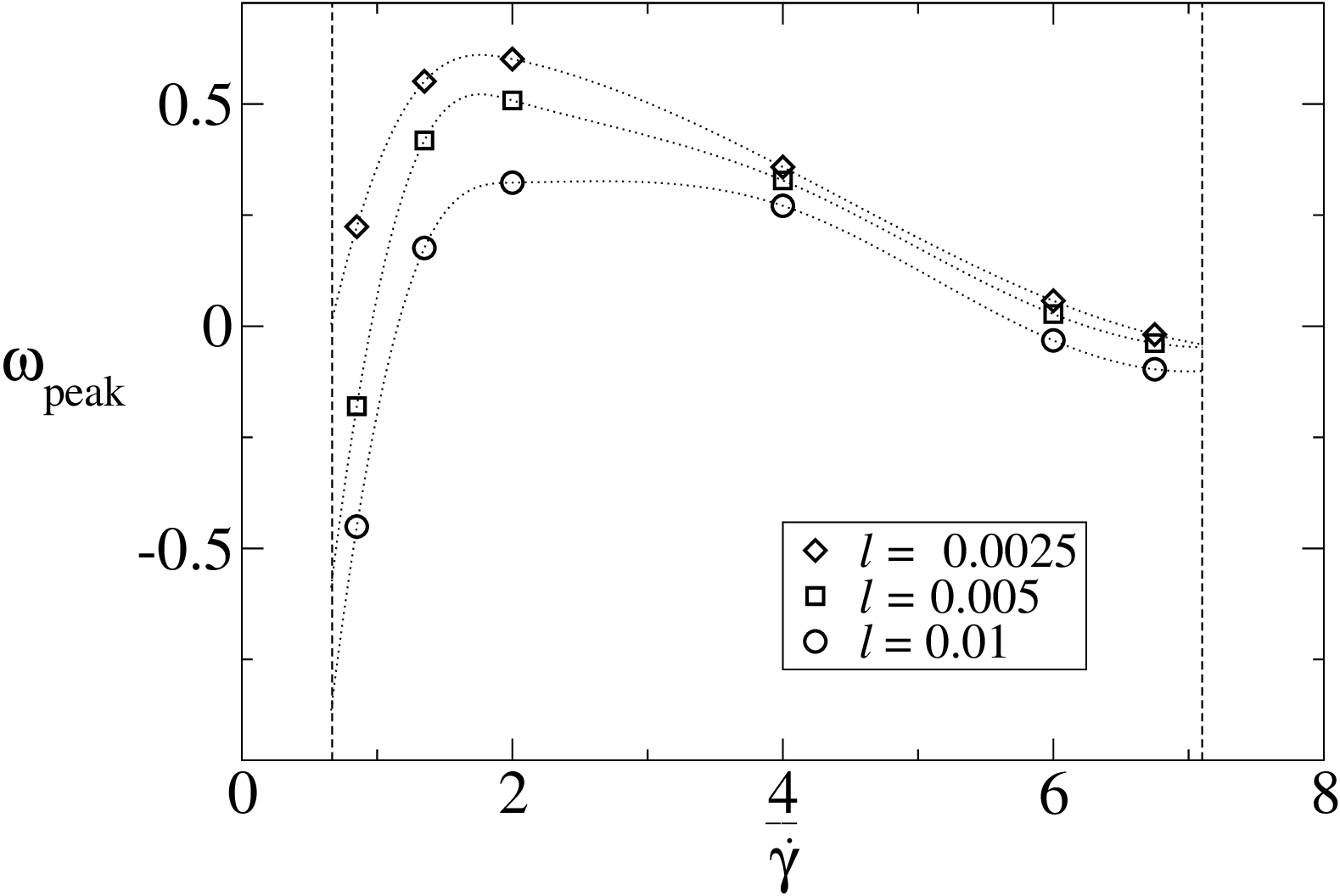}
\caption{Peak of the dispersion relation, {\it i.e.} $\Re \ommax$ at $d\Re\ommax/dq_x=0$.  Parameters as for Fig.~\ref{fig:maxEigen}.  Limits of the banding regime shown by vertical lines. Symbols: data. Dotted lines: cubic splines, as a guide to the eye.
\label{fig:maxima} } 
\end{figure}
%

%
%
\begin{figure}[ht]
\includegraphics[scale=0.5]{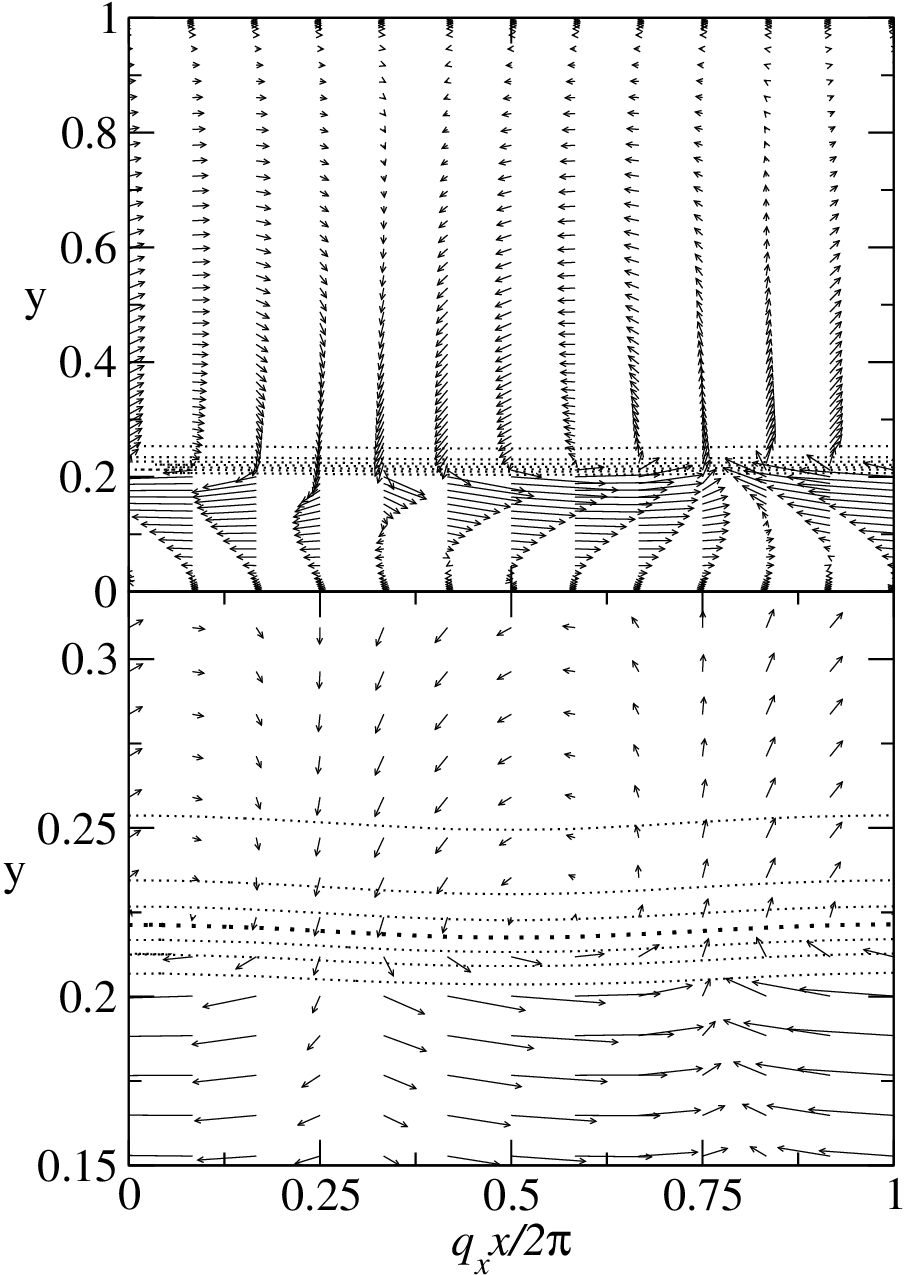}
\caption{ Perturbation to  flow field $s_1\Re\vecv{v}(y)e^{iq_xx}$
  (arrows), and contour lines of perturbed normal stress
  $\tilde{\Sigma}_{xx}(x,y)=\Sigma_{xx}(y)+s_2\Re
  \sigma_{xx}(y)e^{iq_xx}$ (dotted lines), corresponding to the
  eigenvalue of Fig.~\ref{fig:maxEigen} with $l=0.01$, $q_x=2.0$.
  Contours downwards: 0.45, 0.60, 0.75, 0.90, 1.05, 1.20, 1.35 (middle
  value shown thicker). Arbitrary scale factors $s_1=1.5$ and
  $s_2=0.3$.
\label{fig:eigenVector1} } 
\end{figure}
%

For any base profile $\vecv{\Phi}(y)$ and wavevector $\vecv{q}$, the
number of eigenmodes is equal to the number of order parameters summed
over all mesh points. In this Letter, we only consider the eigenvalue
$\ommax(\vecv{q})$ with the largest real part,
$\Re\,\ommax(\vecv{q})$. In particular, we ask if this mode is stable,
$\Re\,\ommax<0$, or unstable, $\Re\,\ommax>0$.  All results given are
for a low solvent viscosity $\eta=0.05\ll G\tau\equiv 1$, consistent
with experiment. We set $a=0.3$, although our findings are
qualitatively robust to variations in $a$. This leaves the applied
shear rate $\gdotbar$ as the tunable parameter.

The dispersion relation $\Re\,\ommax(q_x, q_z=0)$ for fluctuations
with wavevector confined to the direction of the unperturbed flow is
shown in Fig.~\ref{fig:maxEigen} for $\gdotbar=2.0$.  At any $q_x$,
$\Re\ommax$ increases with decreasing $l$, and for small enough $l$
the dispersion relation is positive over a range of wavevectors,
rendering the 1D profile unstable. For small $l$ this applies to shear
rates right across the stress plateau of Fig.~\ref{fig:flowCurve}, as
shown in Fig.~\ref{fig:maxima}.  Because the $l$ values accessed here
-- $l=O(1-10\mu {\rm m})$ for a 1mm rheometer gap -- are even larger than
those expected physically, $l=O(100{\rm nm})$, our results suggest
that, experimentally, the entire stress plateau will be unstable.

In the limit $l\to 0$, $q_x\to 0$, the corresponding eigenfunction
$\{\partial_y v_x, v_y=0, \sigma_{\alpha\beta}(y)\}$ tends to the
spatial derivative of the base state, $\partial_y\{\partial_y V_x,
V_y=0, \Sigma_{\alpha\beta}\}$, representing a simple displacement of
the interface in the flow-gradient direction, with small corrections
in the bulk phases to maintain $\gdotbar={\rm constant}$. As $q_x$
increases from zero, this displacement is modulated by a wave of
wavevector $q_x\hat{\vecv{x}}$ with an eigenvalue $\ommax(q_x) =
\omega_0 + iq_x \omega_1 + q_x^2 \omega_2$ with $\omega_2>0$,
signifying instability. A natural question is whether this instability
has the same origin as that described by McLeish for the local
model~\cite{Mcleish87}. It is not obvious, a priori, that this should
be true because, for the base state at least, the limit $l\to 0$ is
singular~\cite{lu99}. Indeed, a detailed analysis (work in progress)
is more complicated in this case, and deferred to a longer
publication. However, the numerical results of
Fig.~\ref{fig:eigenVector1} are qualitatively consistent with the
mechanism of McLeish, as follows. A wavelike interfacial displacement
with extrema at $q_x x/2\pi = 0.0, 0.5, 1.0$ causes an interfacial
tilt near $q_xx/2\pi=0.25, 0.75$, exposing the normal stress jump
$\Delta\Sigma_{xx}$ across the interface (recall
Fig.\ref{fig:profile}).  This triggers a horizontal perturbation to
the flow field $\Im v_x$ in these regions, which recirculates, giving
an $O(q_x^2)$ vertical velocity $\Re v_y$ at $q_x x/2\pi = 0.0, 0.5,
1.0$. This enhances the original displacement and so causes
instability.  Stability is restored for higher $q_x$
(Fig.~\ref{fig:maxEigen}), a feature that is absent in the local case.

\begin{figure}[t]
\vspace{0.5cm}
\includegraphics[scale=0.30]{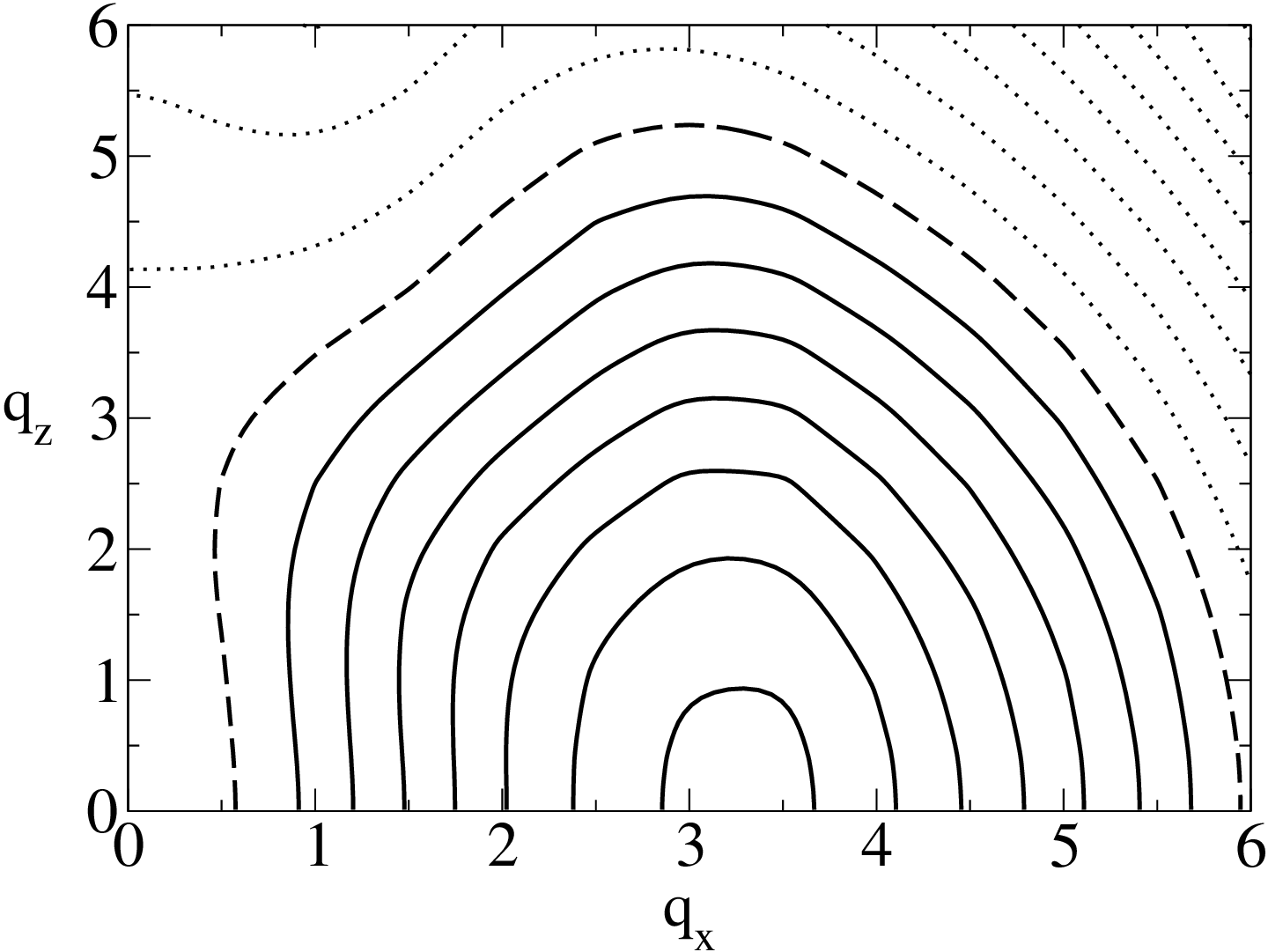}
\caption{Real part of the most unstable eigenvalue.  $a=0.3$,  $\eta=0.05$,  $\gdotbar=2.0$, Reynolds number $\rho/\eta=0.01$ (negligible),  $l=0.01$. Contours are $-0.45, -0.40 ...$ (dotted line), $0.00$ (dashed) and $... 0.25, 0.30$  (solid).
\label{fig:xzPlane} } 
\end{figure}

The eigenvalue $\Re\,\ommax(\vecv{q})$ over the $(q_x,q_z)$ plane is
shown in Fig.~\ref{fig:xzPlane}. 
Modes with wavevector along the $q_x$ axis are much more prone to
instability than those along the $q_z$ axis.  Nonetheless, for smaller
values of $l$ (not shown), modes along the line $q_x=0$ can go
unstable as well.

We note finally an important bound on the validity of our calculation.
The expansion used to obtain Eqn.~\ref{eqn:eigen} is valid for
perturbations that are small at any point in space. For example, for
the stress components we require $\sigma_{\alpha\beta} \ll 1$.
Displacement of the interface by a distance $\Delta$ gives
$\sigma_{\alpha\beta}=\Delta\, d\Sigma_{\alpha\beta}/dy$, which is
$O(\Delta/l)$, because the base profile $\Sigma_{\alpha\beta}$ changes
by $O(1)$ over the interfacial width $O(l)$. We are thus restricted to
small displacements, $\Delta\ll l$.  In future work, we will consider
$\Delta\gg l$.

We comment briefly on the stability of a sheared interface between two
phases of a binary fluid in ``model H''~\cite{hohenberg77}. Although
this was studied in Ref.~\cite{BraCavTra02}, that work integrated over
space to get a simple equation for the position of the interface. Such
an approach neglects changes in the interface's profile, and any fluid
flow normal to the interface, so is not guaranteed to agree with ours.
Nonetheless, we found the interface to be stable, as in
Ref.~\cite{BraCavTra02}.  This supports the idea that normal stresses
(absent in model H) cause the instability described above.

To conclude, we have found 1D planar shear banded flow to be linearly
unstable to fluctuations with wavevector in the plane of the banding
interface, within the non local Johnson Segalman model. This applies
to shear rates right across the stress plateau, suggesting that the
instability is ubiquitous and therefore that the existing theoretical
picture of two stable shear bands separated by a steady interface
needs further thought.  Indeed, our finding is consistent with
accumulating evidence for erratic
fluctuations~\cite{WunColLenArnRou01}
and band breakup~\cite{lerouge_thesis} in several systems.  Future
work will study the fate of the interface in the non linear regime,
beyond the validity of this linear study. One possibility is that the
instability is self limiting beyond a critical amplitude set by $l$
({\it e.g.}, $l^{1/2}$). This would be consistent with a narrowly
localized but still unsteady interface, which might be interpreted as
steady in experiments that did not have high spatial resolution. This
might even reconcile early data showing apparently steady interfaces
with recent work revealing fluctuations. 

By contrast, if the instability were found not to be self limiting,
and yet ubiquitous in existing banding models (work in progress), one
would then need a new theoretical picture of (reasonably) steady shear
bands that could still accommodate the required normal stress jump
across the interface. Other open questions include the status of the
instability in curved Couette geometry; and the relative importance of
instabilities at non-zero $\vecv{q}$ (as studied here) to those found
at zero $\vecv{q}$ in recent models of spatio-temporal
rheochaos~\cite{FieOlm04}.

%

The author thanks Paul Callaghan, Mike Cates, Tanniemola Liverpool,
Tom McLeish, Peter Olmsted and Helen Wilson for useful discussions and
feedback; and the EPSRC GR/S29560/01 for funding.


\vspace{-0.3cm}


\end{document}